%% file: main.tex
\documentclass[conference]{IEEEtran}
\IEEEoverridecommandlockouts
\usepackage{cite}
\usepackage{amsmath,amssymb,amsfonts}
\usepackage{graphicx}
\usepackage{textcomp}
\usepackage{xcolor}
\def\BibTeX{{\rm B\kern-.05em{\sc i\kern-.025em b}\kern-.08em
    T\kern-.1667em\lower.7ex\hbox{E}\kern-.125emX}}
    
\usepackage{graphicx}
\usepackage{algorithm}
\usepackage[noend]{algpseudocode}
\usepackage{xcolor}
\usepackage{listings}
\usepackage[most]{tcolorbox}

\usepackage{url}

\expandafter\def\expandafter\UrlBreaks\expandafter{\UrlBreaks
  \do\a\do\b\do\c\do\d\do\e\do\f\do\g\do\h\do\i\do\j%
  \do\k\do\l\do\m\do\n\do\o\do\p\do\q\do\r\do\s\do\t%
  \do\u\do\v\do\w\do\x\do\y\do\z\do\A\do\B\do\C\do\D%
  \do\E\do\F\do\G\do\H\do\I\do\J\do\K\do\L\do\M\do\N%
  \do\O\do\P\do\Q\do\R\do\S\do\T\do\U\do\V\do\W\do\X%
  \do\Y\do\Z}

\definecolor{commentgreen}{rgb}{0.3,0.5,0.5}
\definecolor{keyred}{rgb}{0.63,0.129,0.258}
\definecolor{codegray}{rgb}{0.1,0.1,0.1}
\definecolor{codepurple}{rgb}{0.58,0.4,0.82}
\definecolor{backcolour}{rgb}{0.95,0.95,0.92}
\definecolor{maroon}{rgb}{0.87,0.68,0.32}

\usepackage{soul}

\newcommand{\add}[1]{\textcolor{blue}{#1}}

\lstdefinelanguage{JavaScript}{
  keywords={typeof, new, true, false, catch, return, null, catch, switch, var, if, in, while, do, else, case, break, require, uint, assert, address},
  keywordstyle=\color{blue}\bfseries,
  ndkeywords={class, export, boolean, throw, implements, import, this, function},
  ndkeywordstyle=\color{darkgray}\bfseries,
  identifierstyle=\color{black},
  sensitive=false,
  comment=[l]{//},
  morecomment=[s]{/*}{*/},
  commentstyle=\color{purple}\ttfamily,
  stringstyle=\color{red}\ttfamily,
  morestring=[b]',
  morestring=[b]"
}

\lstdefinestyle{mystyle}{
	language=JavaScript,   
    commentstyle=\itshape\color{commentgreen},
    keywordstyle=\bfseries\color{keyred},
    numberstyle=\small,
    stringstyle=\color{codepurple},
	basicstyle=\small,
    breakatwhitespace=false,         
    breaklines=true,                 
    captionpos=b,                    
    keepspaces=true,                 
    numbers=left,                    
    numbersep=2pt,                  
    showspaces=false,                
    showstringspaces=false,
    showtabs=false,                  
    tabsize=2,
    frame={bottomline}
}
\usepackage{xspace}
\newcommand{\name}{DEPOSafe\xspace}

\begin{document}

\title{DEPOSafe: Demystifying the Fake Deposit Vulnerability in Ethereum Smart Contracts}

\author{
\IEEEauthorblockN{Ru Ji$^{1*}$, Ningyu He$^{2*}$\thanks{*The first two authors contributed equally to this work.}, Lei Wu$^{3}$, Haoyu Wang$^{1}$, Guangdong Bai$^{4}$, Yao Guo$^{2}$}
\IEEEauthorblockA{
$^{1}$ Beijing University of Posts and Telecommunications, Beijing, China\\
$^{2}$ Peking University, Beijing, China	
$^{3}$ Zhejiang University, Hangzhou, China	\\
$^{4}$ The University of Queensland, Australia	\\
}
}
\maketitle

\begin{abstract}
Cryptocurrency has seen an explosive growth in recent years, thanks to the evolvement of blockchain technology and its economic ecosystem. Besides Bitcoin, thousands of cryptocurrencies have been distributed on blockchains, while hundreds of cryptocurrency exchanges are emerging to facilitate the trading of digital assets. 
At the same time, it also attracts the attentions of attackers.
\textit{Fake deposit}, as one of the most representative attacks (vulnerabilities) related to \textit{exchanges} and \textit{tokens}, has been frequently observed in the blockchain ecosystem, causing large financial losses. 
However, besides a few security reports, our community lacks of the understanding of this vulnerability, for example its \textit{scale} and the \textit{impacts}.
In this paper, we take the first step to demystify the fake deposit vulnerability. Based on the essential patterns we have summarized, we implement {\name}, an automated tool to detect and verify (exploit) the fake deposit vulnerability in ERC-20 smart contracts. {\name} incorporates several key techniques including \textit{symbolic execution based static analysis} and \textit{behavior modeling based dynamic verification}.
By applying {\name} to 176,000 ERC-20 smart contracts, we have identified over 7,000 vulnerable contracts that may suffer from two types of attacks. Our findings demonstrate the urgency to identify and prevent the fake deposit vulnerability.
\end{abstract}


\input{intro.tex}

\input{background.tex}

\input{vulnerability.tex}
\input{approach.tex}

\input{evaluation.tex}
\input{discussion.tex}

\input{related.tex}
\input{conclusion.tex}

\bibliographystyle{IEEEtran}
\bibliography{citation.bib}

\end{document}

%% file: intro.tex
\section{Introduction}
\label{sec:intro}

As the first generation blockchain platform, Bitcoin demonstrated that it is possible to use the Internet to construct a decentralized value-transfer system that can be shared across the world and virtually free to use.
Due to performance and scalability issues, however, it is difficult (if not impossible) for Bitcoin to support complex applications.
To this end, Ethereum~\cite{ethereum} was proposed to allow users to create \textit{DApp}s (Decentralized Applications) by developing \textit{smart contract}s, which have been regarded as the most creative blockchain technique after Bitcoin. 

Up to April 2020, there exist more than 25 million smart contracts in Ethereum. However, only 0.36\% of them have released their source code according to our dataset, which reflects the dilemma between security and privacy.
Besides, Ethereum uses \textit{Ether} as its official cryptocurrency, which can be held and transferred between \textit{account}s in the Ethereum network.
Besides Ether, Ethereum also allows users to create and issue cryptocurrencies (\textit{token}s). 
To standardize the token behaviors, Ethereum proposes a technical standard named ERC-20~\cite{erc20-guide} to ensure the interoperability between tokens.
As a result, Ethereum has attracted lots of attention and become one of the most representative second generation blockchain platforms. 
By the end of April 2020, there are over 176,000 ERC-20 tokens on the Ethereum platform.

With the rapid development of Ethereum, \textit{cryptocurrency exchange} (\textit{exchange} for short), as the third-party supporting service, has emerged to support \textit{fiat-to-crypto} and \textit{crypto-to-crypto} trades.
According to their trading models, exchanges can be categorized into two types, \textit{centralized exchange} (CEX) and \textit{decentralized exchange} (DEX).
CEX, as the name suggests, requires a central entity as the intermediary to complete token transfer between its users. Therefore, the trustworthiness of the middle man plays an important role in this transaction model.
Contrary to CEX, DEX removes the entity in the middle to store data, and performs token exchange with a matchmaking tradeoff model~\cite{dex}. 
Theoretically, a DEX is composed of smart contract(s) without the need of human intervention. 
Therefore, DEX has been developed to allow peer-to-peer trading of cryptocurrencies without an intermediary but solely depends on smart contracts.
Such a trading model can not only guarantee the privacy of users\footnote{Although it is consistent with the anonymity of blockchain, abandoning the Know Your Customer (KYC)~\cite{kyc} policy is still a concern.}, but also ensure the behaviors of transaction strictly follow the logic encoded in smart contracts.

Unfortunately, a number of security issues have been found in tokens and exchanges, which attracted attacker attention as well. First, a number of attacks targeting exchanges have been observed in the wild~\cite{exchange-attack-1,exchange-attack-2,exchange-attack-3}, which caused great financial losses.
Besides, ERC-20 related bugs have been underestimated by developers~\cite{chendefining}. For example, a token may be accidentally created as non-standard without fully understanding the ERC-20 standard.
As a result, \textbf{the combination of the \textit{flawed verification} of exchanges and \textit{non-standard implementation} of tokens might cause severe damages.}

\textit{Fake deposit} is one of the most representative vulnerabilities related to both exchanges and tokens. 
It is publicly well known as it has appeared in recent security reports and news~\cite{fakedeposit-attack-1,fakedeposit-attack-2,fakedeposit-attack-3,fakedeposit-attack-4}. 
The attacks exploiting this vulnerability have remained active for a long time, observed in Ethereum and other platforms (e.g., USDT and EOSIO).
Unlike other vulnerabilities, the fake deposit vulnerability is conditioned by two requirements, i.e., the non-standard implementation of a token and the flawed verification of an exchange.
To be specific, \textit{deposit} means a user transfers a certain type of token into the exchange. A malicious user, however, could take advantage of the flaws in smart contracts of that token and deficient exchange verification mechanism to achieve a \textit{fake deposit}, in which the amount transferred is typically too large to be affordable for the attacker. Consequently, a malicious user can deceive the exchange and gain huge profit with nearly no cost.
To date, the details and impact of this vulnerability have not been well studied.

\textbf{This Paper.}
We take the first step to demystify the \textit{fake deposit} vulnerability in Ethereum and measure the potential security impacts by analyzing a large number of Ethereum ERC-20 smart contracts. We have analyzed over 176,000 ERC-20 contracts deployed between July 2015 to March 2020.
Specifically, we first characterized the vulnerability and identified the essential patterns. 
We then implemented {\name}, an automated tool to identify and verify (exploit) the \textit{fake deposit} vulnerability.
{\name} consists of two major components.
The first component is a static detector implemented based on \textit{Mythril}~\cite{mythril}, an open-source symbolic execution engine towards Ethereum to efficiently perform the detection. 
After that, to eliminate the false positives introduced by the static analysis, we built a dynamic validator 
to emulate the behaviors between token and exchange when users deposit tokens into exchange, in order to confirm the existence of the \textit{fake deposit} vulnerability.
To facilitate our analysis, we categorized these attacks into two types according to the exchanges, i.e., attacks targeting CEX as \textit{Type-I} and attacks targeting DEX as \textit{Type-II}, respectively.
After applying {\name} to over 176,000 ERC-20 smart contracts we collected, we have identified 56 and 7,716 smart contracts that could be exploited by Type-I and Type-II attacks, respectively. 
The technical details of the flawed implementation of token and exchange are presented in \S\ref{sec:vuls}; the attack emulations for both types of attacks are illustrated in \S\ref{sec:approach}.

\textbf{Contributions}
In summary, this paper makes the following main research contributions.
\begin{itemize}
    \item We proposed {\name}, the first framework aims to 1) identify if the smart contract is compliant with the ERC-20 standard rules and subject to fake deposit vulnerability; 2) automatically exploit vulnerable ERC-20 smart contracts and prove the existence of loopholes. 
    \item 
    By applying {\name} to over 176K ERC-20 smart contracts, we have identified more than 7,000 ERC-20 smart contracts with fake deposit vulnerability, which can be exploited by the Type-I or Type-II attacks.
    Our investigation shows that the top 10 most influential but vulnerable ERC-20 tokens account for 984 million USD capitalization, while the affected DEXes still have millions USD volumes per day.
    \item We reported various findings based on the analysis of collected data and further propose mitigating mechanisms for this vulnerability. Moreover, the vulnerability dataset will be released to the community to encourage further study.
\end{itemize}

%% file: background.tex
\section{Background}
\label{sec:background}
In this section, we will briefly introduce the necessary background
knowledge of Ethereum~\cite{ethereum} to facilitate the understanding of our further analyses.

\subsection{Ethereum Account and Transaction}
\label{account-and-transaction}
In Ethereum, \textit{account} is the basic unit to identify an entity in the network. An account is identified by a fixed-length hash-like address. Additionally, the account in Ethereum can be divided into two types: \textit{External Owned Account} (EOA) and \textit{Contract Account}.
An EOA is an ordinary account who can transfer tokens, invoke deployed smart contracts and store received tokens. Moreover, an EOA can deploy a smart contract into a Contract Account\footnote{A smart contract can also deploy another smart contract.}. Specifically, an account can deploy a smart contract by sending a transaction that contain the bytecode of smart contract to address \texttt{0x0}.
After that, all accounts in the Ethereum network are able to invoke the smart contract residing in the Contract Account.

Accounts or smart contracts can interact with each other by sending a \textit{transaction}, which consists of a bunch of data that will be parsed and executed by the target smart contract. 
To be specific, the data consists of the signature of designated function and its required arguments. Notice that, to minimize the size of the transaction transferred, Ethereum matches functions by its first four bytes of the signature that is calculated by the Keccak256 hash function~\cite{keccak256}. For example, the function \texttt{transfer(address,uint256)} will be identified by \texttt{0xa9059cbb}.
Once the target address successfully handles the received transaction, the transaction will be recorded online soon and cannot be erased or modified. However, if something goes wrong within executing a transaction, all of the modifications resulted from this transaction will be reverted.

\subsection{Ethereum Virtual Machine (EVM)}
\label{sec:bac:evm}
EVM is a simple stack-based architecture, and the function of the EVM stack is to store the results of intermittent execution of bytecode instructions (\textit{opcode}s). All the operands of opcodes and the calculation result are pushed onto the \textit{stack}, a basic data structure in EVM. 
In a nutshell, EVM can provide a runtime environment for smart contracts that have been compiled to bytecode to be executed, manage execution of the transaction and transit the blockchain to its new state.

Except for stack, data in EVM can also be stored in other areas: \textit{memory}, \textit{storage} and \textit{calldata}. 
To be specific, storage is a key-value mapping that persists between function calls and transactions. As the data in storage are recorded on the blockchain, it charges more to create and update entities in storage. 
Moreover, EVM will assign a \textit{slot ID} for each variable stored in storage to identify it. The slot ID is determined at compilation time and strictly based on the variables order in the contract code. 
In contrast, retrieving and inserting data from memory is much more cheaper. The memory area, however, will be erased after the current transaction.
Besides above areas, \textit{calldata} is used to store external calls to functions, and it is a read-only byte-addressable space.

\subsection{Smart Contract and Bytecode}
\label{sec:back:smartcontract}
A smart contract is a collection of code and data that reside in Contract Account. They are typically written in high level languages, such as Solidity~\cite{solidity}. They are used to implement arbitrary rules as well as guarantee to produce the same result for decentralized parties.
Smart contracts exist and execute in bytecode format, which is compiled from source code.

The Ethereum bytecode is made up of 144 opcodes~\cite{eth-opcodes}. Additionally, each opcode is encoded as one byte, and represented in hexadecimal format.
For example, \texttt{SSTORE} is encoded as \texttt{0x55} in EVM.
Opcode takes zero or multiple arguments to achieve its functionality.
As we mentioned in \S\ref{sec:intro}, only 0.36\% of contracts have opened up the source code. In order to cover all these contracts, we decide to implement {\name} as a bytecode-level analyzer (see Section~\ref{sec:approach}).

Additionally, the bytecode of smart contract is composed of three parts: \textit{creation code}, \textit{runtime code}, and \textit{swarm code}. 
Creation code includes constructor logic and constructor parameters of a contract. It is generated when the bytecode is compiled and it will be executed only once at the time of deployment.
Runtime code will eventually be stored on-chain. It details the logic and behavior of each function in contract. However, it does not contain any of the constructor parameters.
Swarm code is a little bit different. It does not have any practical meaning and can not be executed by EVM. The swarm code is only a string of hash that is calculated by metadata of current smart contract to index it in database.

\begin{lstlisting}[caption={Standard \texttt{transfer} and \texttt{transferFrom} in ERC-20 token}, label={lst:transfer-example}]
function transfer(address to, uint tokens) public returns (bool success) {
  balances[msg.sender] = safeSub(balances[msg.sender], tokens);
  balances[to] = safeAdd(balances[to],tokens);
  emit Transfer(msg.sender, to, tokens);
  return true; }

function transferFrom(address from, address to, uint tokens) public returns (bool success) {
  // update balances and allowances
  emit Transfer(from, to, tokens);
  return true; }
\end{lstlisting}

\subsection{ERC-20 Standard}
\label{sec:background:erc20}
As one of the most popular and well-known technical standards in Ethereum, \textit{ERC-20} specifies six mandatory functions (\texttt{totalSupply}, \texttt{balanceOf}, \texttt{transfer}, \texttt{transferFrom}, \texttt{approve}, and \texttt{allowance}) for the benefit of Ethereum developers.
As shown in listing~\ref{lst:transfer-example}, the \texttt{transfer} uses function \texttt{safeAdd} and \texttt{safeSub} to update the balance table. If the invoker does not have enough token, they will throw exceptions and terminate the current transaction. As for \texttt{transferFrom}, except for the update of balance table, it also checks if the caller has enough allowance to transfer such an amount of tokens. Any of these three verification fails, the transfer will be terminated immediately.

\subsection{Exchange and Deposit}
\label{sec:background:exchange}

\begin{figure}[tbp]
\centerline{\includegraphics[width=\columnwidth]{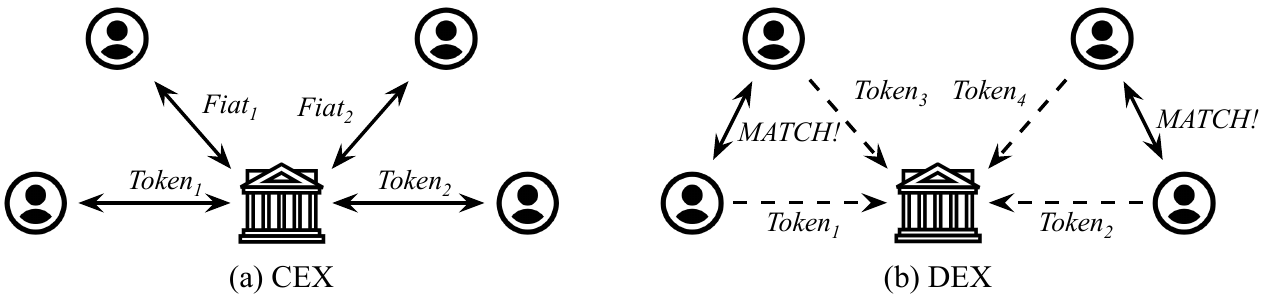}}
\caption{The trading model of CEX and DEX respectively.}
\label{fig:centralized-dex}
\end{figure}

\subsubsection{CEX}
CEX is an intermediary between users in trading process (see Fig.~\ref{fig:centralized-dex}(a)). Typically, the centralized entity is managed by a trustworthy organization or company. However, once the centralized server is down or compromised, it is possible for data breach or even financial loss for both exchange and users. 
Nevertheless, CEX has its advantages: 1) it has ability to achieve quick transactions and support multiple users at once; 2) it supports the exchange between fiat and tokens, or even tokens from different platforms; 3) the trading model determines its scalability and quick response against attacks.

Depositing tokens into CEX is simple. The user invokes a transaction with certain amount of tokens to a designated address. After that, the CEX server will verify if the transaction is successful and update the balance of the user in its database.

\subsubsection{DEX} 
As depicted in Fig.~\ref{fig:centralized-dex}(b), DEX, unlike CEX such as Coinbase~\cite{coinbase} and Binance~\cite{binance}, has no central entity which is managed by a specific company or a person focusing on making a profit.
The DEXes put the control of funds and trading back to users. In other words, DEX does not store user assets, so neither hacker attacks nor the total collapse of the DEX can lead to a loss of funds.
To compete with the traditional CEXes and attract users, these DEXes adopt various business logic. For example, in EtherDelta~\cite{etherdelata}, maker and taker agree on trading off-chain and execute it on-chain, while IDEX~\cite{idex} provides real-time order book updates, which is live on the Ethereum MainNet.
Despite these diverse business logic, DEXes share similar mechanisms, protocols and data models, i.e., the actual trade is executed through a smart contract on blockchain. Therefore, once the transaction is confirmed by the miners, it is impossible to withdraw or cancel it due to the irreversibility. Additionally, all the behaviors through DEX strictly follow the code of its smart contract.

\begin{lstlisting}[caption={DepositToken function}, label={lst:depositToken}]
function depositToken(address token, uint amount) {
  if (msg.value>0 || token==0) throw;
  if (!Token(token).transferFrom(msg.sender, this, amount)) throw;
  tokens[token][msg.sender] = safeAdd(tokens[token][msg.sender], amount);
  Deposit(token, msg.sender, amount, tokens[token][msg.sender]);
}
\end{lstlisting}

To trade tokens through DEX, users have to deposit tokens into smart contract of DEX in advance. Therefore, the user has to firstly grant the permission to DEX to make it eligible; then the user will call the \texttt{depositToken} function which is detailed in listing~\ref{lst:depositToken}.
As we can see, the \texttt{depositToken} takes the token address \textit{token} and deposit value \textit{amount} as arguments, then calls the \texttt{transferFrom} in address \textit{token} to transfer money to DEX's address.

%% file: vulnerability.tex
\section{Fake Deposit Vulnerability}
\label{sec:vuls}

As aforementioned, the attack against \textit{fake deposit vulnerability} depends on both entities of tokens and exchanges.
Therefore, in this section, we will detail the incorrect implementation in ERC-20 token's smart contracts, and also the two types of deficient verification of exchanges. The combination of these two conditions leads to the fake deposit vulnerability.

\subsection{Non-standard Implementation of Tokens}
\label{sec:vul:fake-deposit-vul}
Though ERC-20 enforces the implementation of these interfaces (see \S\ref{sec:background:erc20}), the standard does not specify the implementation details, an incorrect or improper implementation could lead to vulnerable contracts. 
To be specific, the official guideline recommends developers to \texttt{throw} an exception if there are insufficient tokens in the caller's balance to spend, as shown at line 2-3 in listing~\ref{lst:transfer-example}. The \texttt{safeAdd} and \texttt{safeSub} would throw an exception if the overflow happens.

\begin{lstlisting}[caption={An example of a vulnerable implementation of \texttt{transfer}.}, label={lst:false-transfer-example}]
function transfer(address _to, uint256 _value) returns (bool success) {
  if (balanceOf[msg.sender] >= _value && balanceOf[_to] + _value > balanceOf[_to]) {
    balanceOf[msg.sender] -= _value;
    balanceOf[_to] += _value;
    Transfer(msg.sender, _to, _value);
    return true;
  } else { return false; }
}
\end{lstlisting}

However, some developers use a conditional statement to check the caller's balance instead of the assertion statement (see listing~\ref{lst:false-transfer-example}).
If the balance of the caller (identified by \texttt{msg.sender}) is insufficient, the transfer will \texttt{return false} at line 7. Unfortunately, no matter which value returned 
, the current transaction will not be terminated.
\textit{This gap between the actual behavior and the developer's expectation breaks the guideline and leads to the vulnerability.}

\subsection{Flawed Verification of Exchanges}
\label{sec:vul:deficiency-dex}
Except for the vulnerabilities that reside in the ERC-20 token smart contract, successfully exploiting the fake deposit also relies on the implementation flaws in exchanges. Therefore, we then introduce two types of deficiencies which lead to the security threats of exchanges.

\subsubsection{Flawed Token Verification of DEXs}
\label{sec:vul:deficiency-dex:audit}
After a thorough manual inspection towards current (and past) mainstream DEXes, we found that most of them perform deposit logic by calling the function \texttt{depositToken}, which is detailed in listing~\ref{lst:depositToken}.
As line 3 shows, the DEX invokes the traded token's \texttt{transferFrom} (or sometimes \texttt{transfer}) to perform deposit. Notice that the DEX is responsible for auditing the token's smart contract that is traded on its platform to ensure financial security.
However, imagine a token's \texttt{transferFrom} is encapsulated by \texttt{safeTransferFrom}, where it performs security check and then invokes the real \texttt{transferFrom}. 
If the DEX does not audit this smart contract or neglects this unsafe implementation, there seems no security check when DEX calls \texttt{transferFrom} at line 3.

\subsubsection{Flawed Back-end Verification of CEXs}
\label{sec:vul:flawed-cex}
When user deposits tokens into CEXes, the depositing amount is parsed and determined by CEX's back-end server. However, a deficient verification strategy may result in unexpected behaviors.
To be specific, some CEXes do only verify the \texttt{status} field and \texttt{\_value} in the \texttt{input} field of a transaction, which is reported by security analysts~\cite{fakedeposit-attack-chinese}.
The \texttt{status} indicates whether the current transaction is terminated with nothing unusual; the \texttt{\_value} is the value of arguments in the \texttt{transfer} function.
Therefore, if a transaction is terminated by ``return False;'' instead of throwing an exception, the \texttt{status} code will still be \texttt{1}, which means current transaction is terminated normally.
Unfortunately, some CEXes rely on such an insufficient verification to determine how many tokens are deposited. 

%% file: approach.tex
\section{DEPOSafe}
\label{sec:approach}
\begin{figure}[tbp]
\centerline{\includegraphics[width=\columnwidth]{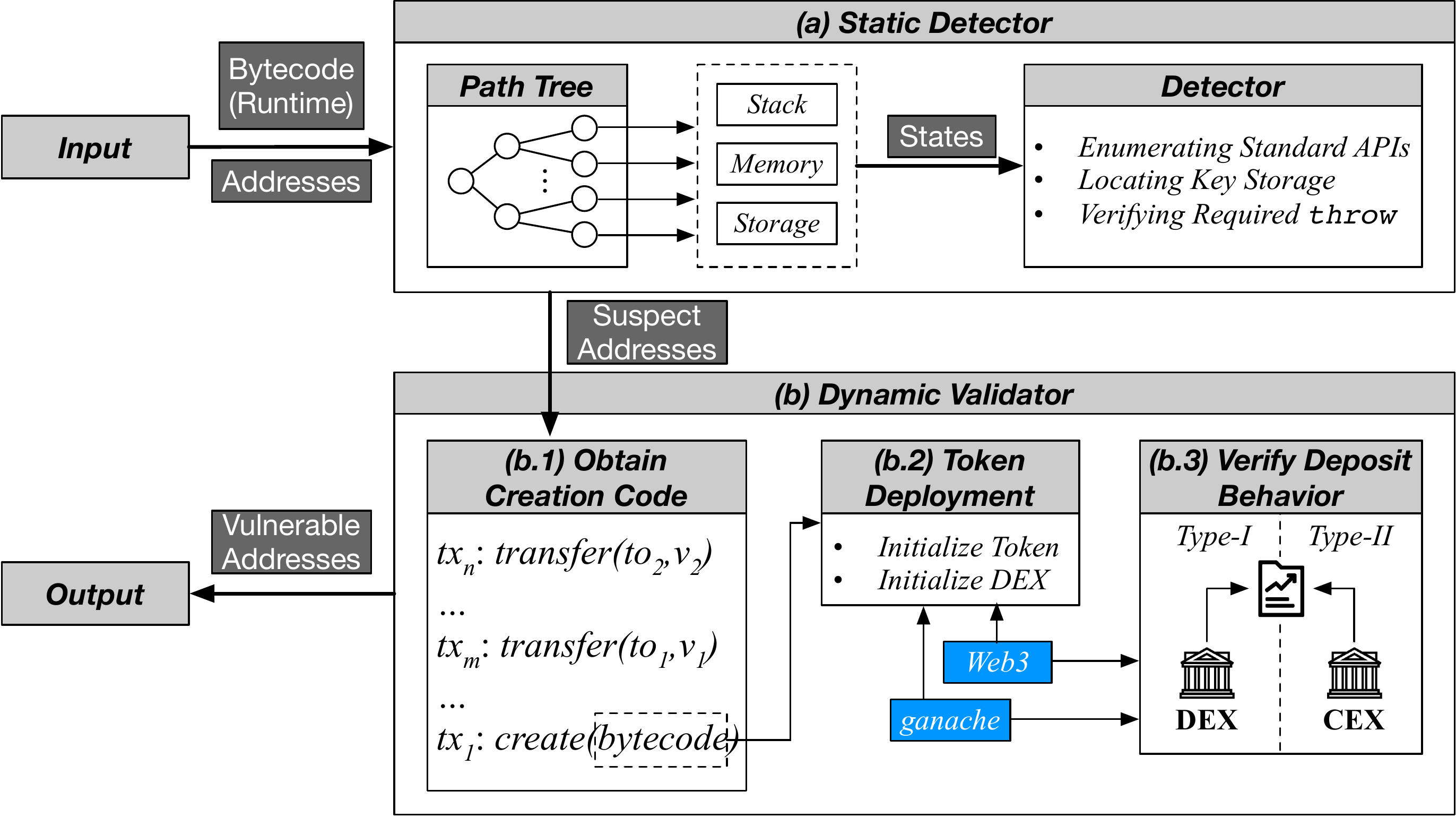}}
\caption{An overview of {\name}.}
\label{fig:framework}
\vspace{-0.2in}
\end{figure}

To evaluate if a smart contract is vulnerable to the fake deposit attack, we implemented {\name}. Fig.~\ref{fig:framework} shows the overall work process of \name. 
It takes the runtime bytecode of smart contracts with their corresponding addresses as input, and generates a security report through a pipeline composed of two parts: \textit{static detector} and \textit{dynamic validator}. 
Specifically, the static detector is implemented based on \textit{Mythril}~\cite{mythril}, in which we screen out the addresses which may be vulnerable to the fake deposit loophole.
Due to the inherent false positives of static analysis, we further implemented in {\name} a dynamic validator to verify the flagged smart contracts are indeed vulnerable to fake deposit.
The dynamic validator takes advantage of \textit{web3}~\cite{web3}, which is a collection of libraries for interacting with a local or remote Ethereum node. We used it to interact with our local private chain provided by \textit{ganache-cli}~\cite{ganache}, which is a customizable blockchain emulator.
We detail both components in the following.

\subsection{Static Detector}
\label{sec:approach:static}

To efficiently detect the fake deposit vulnerability, we propose a static detector relying on symbolic execution to perform the detection.
The Mythril, which we relied on, implements an EVM-like virtual machine, so it is able to imitate the behaviors of a given smart contract on bytecode level. 
Therefore, our detector firstly symbolic executes the bytecode to traverse all the feasible paths. Meanwhile, it records the virtual machine state (including stack, memory, storage, etc.) after each instruction. 
After that, based on the recorded information, the detector is capable of identifying vulnerabilities.

\begin{figure}[tbp]
\centerline{\includegraphics[width=\columnwidth]{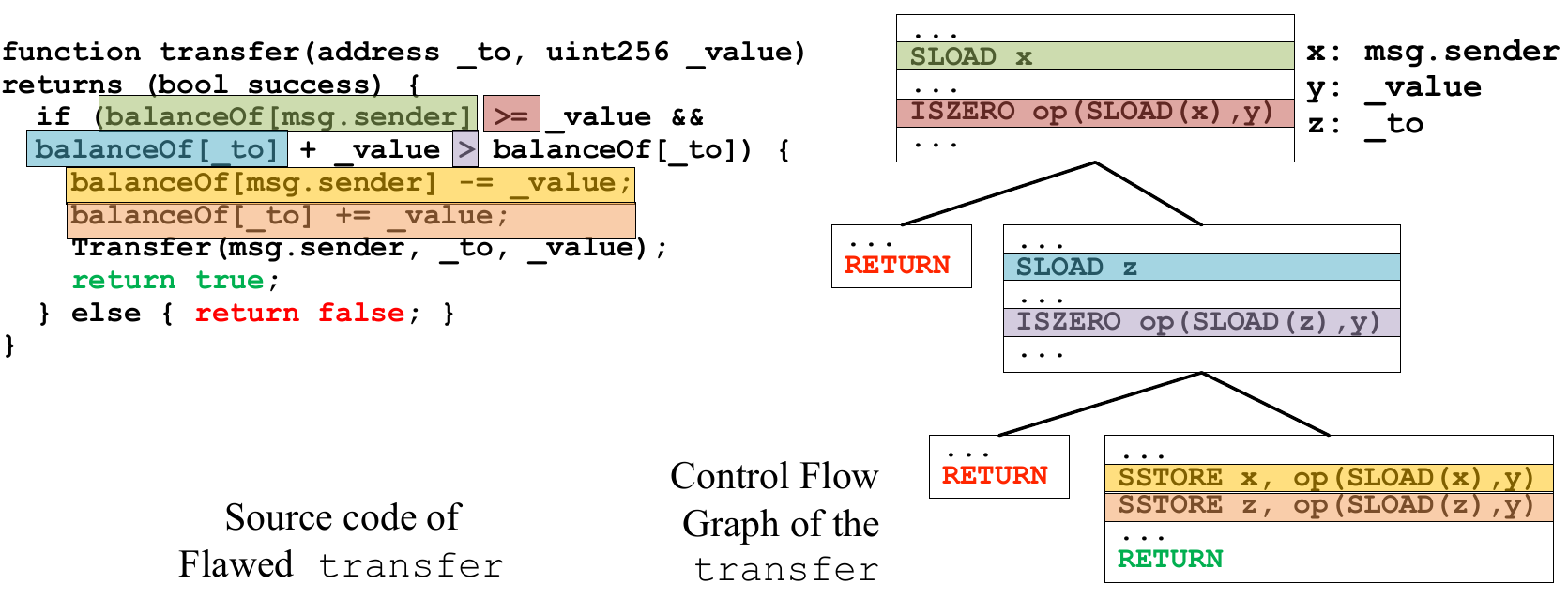}}
\caption{The detection logic of static detector towards a flawed \texttt{transfer}.}
\label{fig:source-bytecode}
\end{figure}

As discussed in \S\ref{sec:vul:fake-deposit-vul}, to accurately detect the incorrect implementation of \texttt{transfer} and \texttt{transferFrom}, we divide the whole static analysis into three steps, including \textit{enumerating standard APIs}, \textit{locating key storage}, and \textit{verifying required \texttt{throw}}.
Specifically, in the first step, we use the signature of functions to find out if the smart contract implements any of the six standard functions; 
in the second step, we use the opcode \texttt{SLOAD} and \texttt{SSTORE} to pinpoint the storage slots of those key arguments, i.e., \texttt{balanceOf} and \texttt{allowance}; 
in the last step, we check the final opcode of relevant paths to identify whether the exception is thrown as required by the ERC-20 standard. 
The detection procedure against the flawed \texttt{transfer} listed in listing~\ref{lst:false-transfer-example} is shown in Fig.~\ref{fig:source-bytecode} and details of these three steps are described in the following.

\subsubsection{Enumerate Standard APIs}
As the existence of \texttt{transfer} and \texttt{transferFrom} is the prerequisite of the fake deposit vulnerability, so firstly we check if the smart contract is compliant with the standard ERC-20 token interface~\cite{erc20-guide}. In other words, the signatures of those six standard functions should be declared in the smart contract's bytecode.
As shown in Fig.~\ref{fig:source-bytecode}, after identifying the existence of \texttt{transfer} (including \texttt{transferFrom} but we use \texttt{transfer} as an example here), we will generate the Control Flow Graph (CFG) and symbolic execute each paths.

\subsubsection{Locate Key Storage}
\label{sec:approach:static:locate}
If the \texttt{transfer} is implemented, we then verify whether it follows the official standard, i.e., \textbf{\texttt{throw}} an exception if there is insufficient tokens in the caller's account. Similarly, we also examine the implementation of \texttt{balanceOf} and \texttt{allowance} in \texttt{transferFrom} as well.
Take function \texttt{transfer} as an example. We have to pinpoint the storage slots used by \texttt{balanceOf} firstly. 
As the Mythril records the virtual machine state when it executes the opcodes of the smart contract, we extract the recorded states of function \texttt{transfer} and traverse all of them to identify the addresses from which the \texttt{balanceOf} is \texttt{SLOAD} (as the balance table is stored in storage (see \S\ref{sec:bac:evm}) and has to be retrieved by instruction \texttt{SLOAD} in EVM).
As we can see from Fig.~\ref{fig:source-bytecode}, the green block indicates where it retrieve\add{s} the balance of \texttt{msg.sender}, i.e., the caller. We mark the label \texttt{x} as \textit{storage address}.
Then, we traverse the storage addresses to examine whether any of them is treated as target address by \texttt{SSTORE} (like \texttt{SLOAD}, EVM has to update variable in storage by instruction \texttt{SSTORE}) in \texttt{transfer}. We can see the yellow block which indicates the balance of \texttt{msg.sender} is updated and stored in storage with label \texttt{x}.
Therefore, the \texttt{balanceOf} in \texttt{transfer} is updated, which could lead to the actual change of balance data.

\subsubsection{Verify Required \texttt{throw}}
\label{sec:approach:static:throw}
Similarly, the success of second step is the precondition of this step. In this step, we detect if the \texttt{balanceOf} is protected and properly handled.
Mythril generates all the feasible paths of the \texttt{transfer} function (if it exists), so we screen out all the instructions that indicates a normal termination of current path, i.e., \texttt{STOP} and \texttt{RETURN}, which means these paths are not terminated by assertion. Like the \texttt{RETURN} colored by green and red in Fig.~\ref{fig:source-bytecode}, they all represent the termination of that path.
Then, we will traverse these paths backward to search for the instruction \texttt{ISZERO}, which is used at the comparison between values. For example, from any of the \texttt{RETURN} opcode in Fig.~\ref{fig:source-bytecode}, we can find a \texttt{ISZERO} backward.
As for the two arguments compared by \texttt{ISZERO}, if one of them is the target address of \texttt{SSTORE}, it means the \texttt{balanceOf} is protected by \texttt{if-else} before updating. Hence, we treat this \texttt{ISZERO} as a \textit{protected node}, as the red block and purple block in Fig.~\ref{fig:source-bytecode}. 
For all the paths forking from the protected nodes, if none of them is terminated by revert instructions, i.e., \texttt{REVERT} and \texttt{ASSERT\_FAIL}, we can ensure that this smart contract does not follow the official guideline. In other words, neither of branches forked by \texttt{if-else} is terminated by assertion, such that the smart contract may be subject to to fake deposit vulnerability.

\smallskip
\smallskip \noindent
Though we have proposed a well-designed strategy to identify the fake deposit vulnerability in ERC-20 smart contracts, we have to face the introduced false positives due to the inherent limitations of static analysis methods.
Taking Listing~\ref{lst:transfer-example} as an example, the smart contact may implement security check by \texttt{safeAdd} or \texttt{safeSub}. This may introduce false positives to {\name}. 
In \S\ref{sec:evaluation:effectiveness}, we comprehensively analyze all the false positives reported by {\name}.
Moreover, we implement a dynamic validator after static detector. This benefits {\name} in two aspects: 1) eliminating the false positives introduced by static analysis and increasing the precision; 2) achieving automatic EXP generation towards fake deposit to validate the existence of vulnerability.

\subsection{Dynamic Validator}
\label{sec:approach:dynamic}
The strategy in our dynamic validator is to mimic the deposit behavior between tokens and exchanges. To this end, we divide the whole validation process into three steps: \textit{obtaining creation code}, \textit{preparing deployment environment}, and \textit{verifying deposit behaviors}.

\subsubsection{Obtain Creation Code}
As the static detector only passes the addresses of those token smart contracts that may be vulnerable to fake deposit behaviors, we have to firstly obtain the creation code, which indicates the deployed contract and the corresponding initial value of parameters (see \S\ref{sec:back:smartcontract}).
To do this, we firstly implement a crawler to grab all the transactions invoked and received from \textit{etherscan.io}~\cite{etherscan} (a well-known and credible browser for Ethereum) according to the addresses.
Then, we only keep the oldest one, which is the contract deployment transaction (see \S\ref{account-and-transaction}), to parse.
Finally, we can obtain the creation code and the initial values from the \texttt{input data} field. 

\subsubsection{Prepare Deployment Environment}
\begin{figure}[tbp]
\centerline{\includegraphics[width=1\columnwidth]{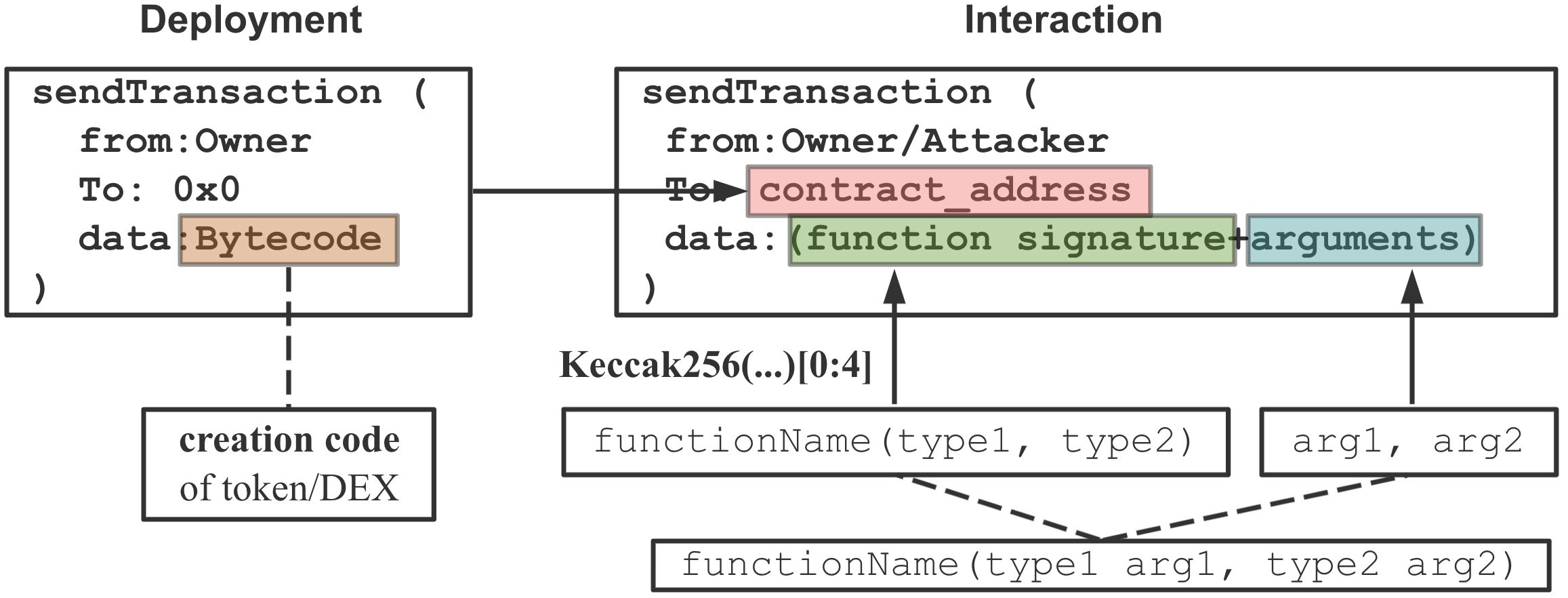}}
\caption{The deployment and interaction between contracts in favor of web3.}
\label{fig:sendtransaction-explain}
\end{figure}

After getting the creation code of tokens, the next step is to deploy them to perform test. To avoid ethical and financial issues, we conduct the testing on our private chain.
To be specific, we set up a private chain and interact with the deployed contract in favor of \textit{ganache} and \textit{web3} respectively. Moreover, as shown in Fig.~\ref{fig:sendtransaction-explain}, the \texttt{sendTransactions} provided by web3 enables the contract deployment and function invocation from us. 
Except for tokens, as DEXes also perform all their functionalities in smart contracts (see \S\ref{sec:background:exchange}), we also deployed the DEXes in our testing environment for the following test.

\subsubsection{Verify Deposit Behaviors}
\label{sec:approach:verifying}
As shown in Fig.~\ref{fig:sendtransaction-explain}, we use the \texttt{sendTransaction} to invoke the functions in deployed smart contracts. In \texttt{sendTransaction}, we can indicate the invoker, receiver, and the specific function and its parameters in \texttt{from}, \texttt{to}, and \texttt{data} respectively.

\begin{figure}[tbp]
\centerline{\includegraphics[width=\columnwidth]{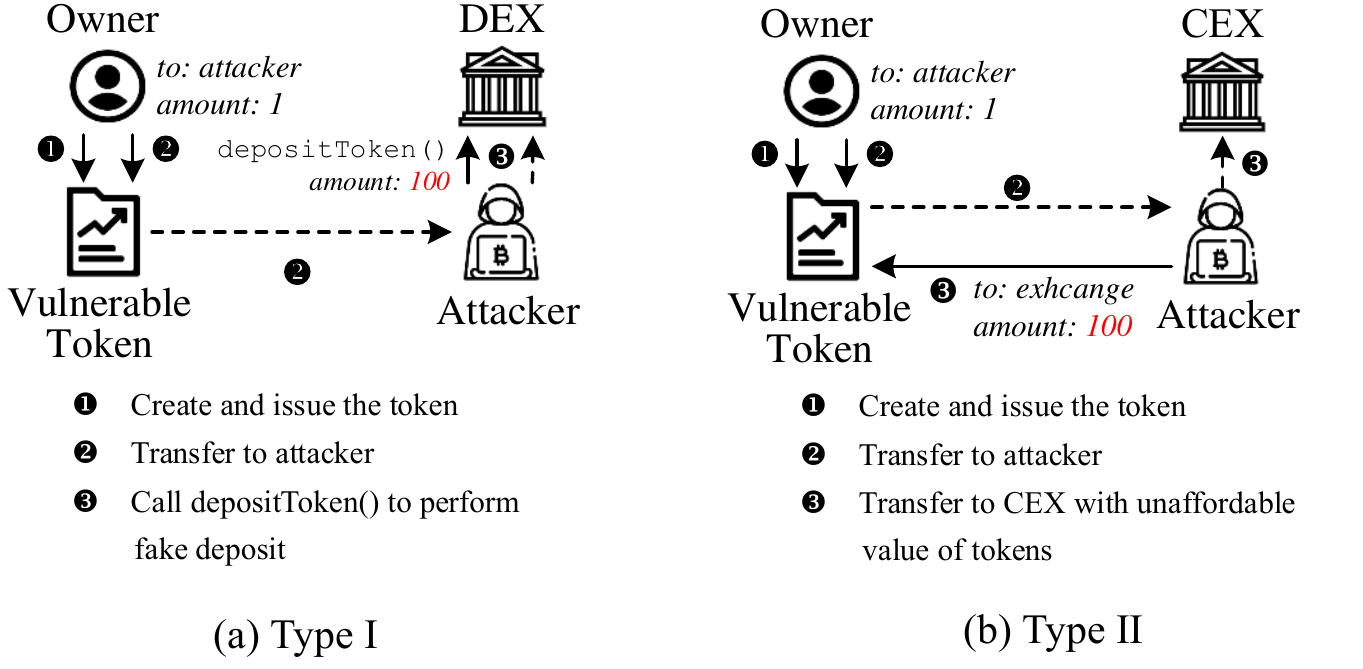}}
\caption{The fake deposit attack resulted from (a) the incorrect implementation of ERC-20 token and lack of auditing from DEX; (b) the insufficient verification of back-end server of CEX. (Where the solid line is the ``Call Flow'' and the dash line is the ``Money Flow'')}
\label{fig:fake-deposit-attack}
\end{figure}

As explained in \S\ref{sec:vul:deficiency-dex}, there are two types of flawed verification of DEX and CEX, which requires different deposit behaviors, i.e., different testing processes. Therefore, we divide the verifying process into two parts (as shown in Fig.~\ref{fig:fake-deposit-attack}), which aim to verify the vulnerabilities existed in DEX and CEX. 

\textbf{Type-I Attack.}
Type-I attack is related to the DEXes. Under normal circumstances, the \texttt{depositToken} in DEXes' smart contract would invoke the traded token's \texttt{transferFrom} or \texttt{transfer}. However, the mis-implementation of functions in ERC-20 tokens would result in Type-I attack.
Except for the example raised in \S\ref{sec:vul:deficiency-dex:audit}, there exists some other cases. If the smart contract does not implement the \texttt{transferFrom} or \texttt{transfer} at all, the invocation from \texttt{depositToken} will be automatically executed by the \textit{fallback} function~\cite{fallback}, which will be executed if none of other functions matches to the name indicated in the incoming function call. 
Consequently, some unexpected behaviors would happen.

Accordingly, we design our verifying procedures for Type-I attack as shown in Fig.~\ref{fig:fake-deposit-attack}(a).
Notice that, we have already deployed the smart contracts of ERC-20 and DEX before this step starts.
Firstly, we (the owner) issue the corresponding tokens according to the ERC-20 contract.
Then, the owner transfers only 1 newly issued token to another account, i.e., the attacker.
Next, the attacker invokes \texttt{depositToken} by calling the DEX to deposit 100 tokens which is definitely unaffordable to the attacker. If the smart contract is implemented incorrectly, i.e., not following the ERC-20 guideline like examples depicted in the last paragraph, the deposit request will be handled.
Finally, we examine the balance table in the DEX to verify the final result of such a deposit request. If the attacker has balance which is equal to the amount of tokens deposited into the DEX, the Type-I fake deposit attack is considered successful.

\textbf{Type-II Attack.}
The Type-II attack relies on the flaws in back-end source code of CEX, as discussed in \S\ref{sec:vul:flawed-cex}. 
Though it is impossible for us to obtain the source code, we should not underestimate the financial losses which may result from this type of attack. 
Therefore, we just assume the CEX has a deficiency in validating transactions and perform attacks following the behaviors detailed in Fig.~\ref{fig:fake-deposit-attack}(b).
Specifically, like the verification of Type-I attack, we firstly issue the token and transfer only 1 token to attacker account.
Then, however, the attacker would call the \texttt{transfer} function in the ERC-20 with the \texttt{\_to} as address of the CEX and \texttt{\_value} as 100, which is also unaffordable to him.
After that, we would examine this transaction in our private chain to check if the \texttt{status} is \texttt{1}. 
If it is, we can assume this kind of token has possibility to be fake deposited into some defective CEXes. Therefore, we have measured the most worst case for the fake deposit behavior between ERC-20 and CEX in Type-II attack.

%% file: evaluation.tex
\section{Evaluation}
\label{sec:evaluation}

Our evaluation is driven by the following research questions:

\begin{itemize}
    \item[RQ1] \textit{How many ERC-20 smart contracts are vulnerable to fake deposit attacks, and can be successfully exploited?} We want to measure the overall landscape of fake deposit vulnerability across all the ERC-20 tokens.
    \item[RQ2] \textit{How effective is {\name} in identifying fake deposit vulnerability?} We want to measure the precision and recall of {\name}.
    \item[RQ3] \textit{What is the impact of the fake deposit vulnerability?} 
\end{itemize}

To answer RQ1, we apply the {\name} to all of the 176k ERC-20 smart contracts deployed by the time of this writing.
To evaluate the effectiveness of {\name} (RQ2), we manually select samples and perform a fine-grained analysis on the false positives and false negatives introduced by both static detector and dynamic validator in {\name} respectively.
To answer RQ3, we further analyze the transactions and volume of these vulnerable ERC-20 tokens and corresponding DEXes, as an indicator to measure the overall impacts.

\subsection{RQ1: Vulnerable ERC-20 Smart Contracts}

\subsubsection{Overall Result}
After analyzing 176,559 ERC-20 smart contracts deployed before April 2020, 7,735 (4.38\%) smart contracts are marked as vulnerable by {\name}.
Among them, 56 smart contracts are vulnerable to Type-I attack (see \S\ref{sec:approach:verifying} and Fig.~\ref{fig:fake-deposit-attack}), while 7,716 of them can be exploited by the Type-II attack\footnote{37 smart contracts can be exploited by both types of attacks.}.
It suggests that \textit{fake deposit vulnerability is prevalent in the ERC-20 smart contract ecosystem}. 

Note that, for the 56 smart contracts that are vulnerable to Type-I attack, they can be indeed exploited by attackers in the wild, due to their vulnerable implementation. 
For the 7,716 smart contracts that are vulnerable to Type-II attack, whether the attacks could be successfully performed are also relying on the verification of CEXes. 
As aforementioned in \S\ref{sec:vuls}, performing Type-II attack also requires the flawed verification of CEXes, including lacking audit to tradable token and deficiency in back-end verification. As we are unable to acquire the source code of CEXes, we can only measure the vulnerability from the perspective of smart contracts.
Nevertheless, these vulnerabilities are indeed existing, which pose great security issues, especially when considering the fake deposit attacks are frequently reported from time to time.

\subsubsection{Distribution of the Vulnerable ERC-20 Smart Contracts}
\begin{figure}[tbp]
\centerline{\includegraphics[width=\columnwidth]{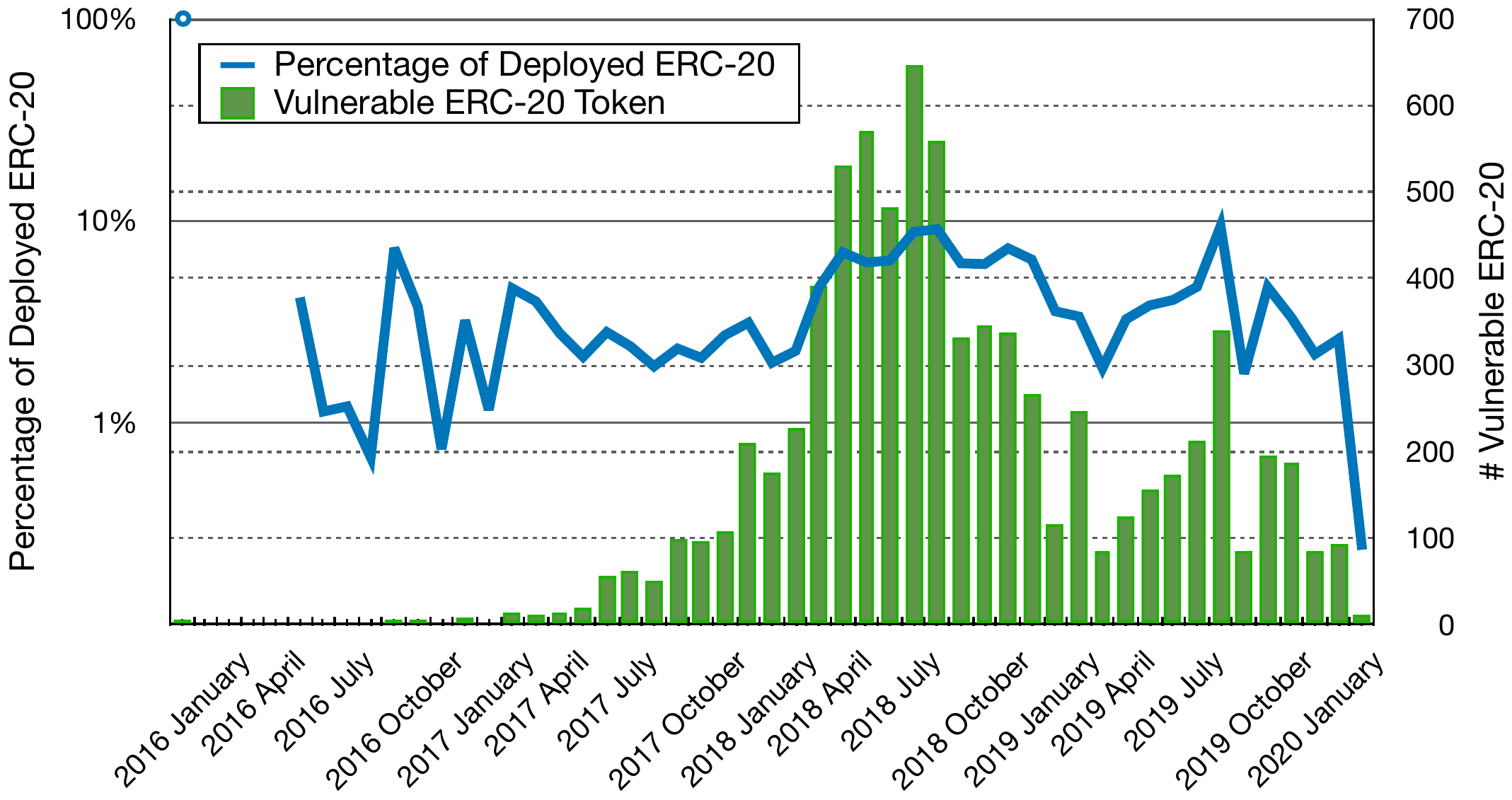}}
\vspace{-0.1in}
\caption{The distribution of ERC-20 smart contracts with fake deposit vulnerability, and its percentage of deployed ones in each month.}
\label{fig:fake-deposit-distribution}
\vspace{-0.2in}
\end{figure}

We further analyze the distribution of these 7,735 vulnerable smart contracts based on their deployment time.
As depicted in Fig.~\ref{fig:fake-deposit-distribution}, the number of smart contracts that are vulnerable to fake deposit vulnerability shows an increasing trend till June 2018, which may be due to the growing popularity of ERC-20 tokens. 
After that, the amount of vulnerable contracts declined greatly. We speculate the decline is related to a mass of attack behaviors towards Ethereum, which in turn reminds the developers of newly deployed ERC-20 tokens. 

Fig.~\ref{fig:fake-deposit-distribution} also shows the proportion of the vulnerable ERC-20 contracts within the deployed ones in each month.
As we can see, for the first month (January, 2016), all of the 5 ERC-20 tokens are affected, which may due to the developer's unfamiliarity to ERC-20 standard. 
For the following four months, all the deployed ERC-20 contracts are not vulnerable to fake deposit vulnerability. It might be that the number of new deployed ERC-20 contracts are small, which are between 15 to 33.
From then on, the percentage fluctuated between 1\% to 10\%, even if the absolute number changes dramatically. It suggests that although attacks related to this vulnerability have been reported from time to time, there are always a considerable number of developers who overlooked it.

\subsection{RQ2: The Effectiveness of {\name}}
\label{sec:evaluation:effectiveness}

{\name} is composed of two components: \textit{static detector} and \textit{dynamic validator}. Due to the inherent limitation of static analysis, the false positives will be introduced inevitably. Conversely, though dynamic validator can ensure the exploited smart contracts are really vulnerable, it will introduce false negatives, as some vulnerable contracts cannot be easily triggered. Therefore, we seek to evaluate the effectiveness of the static detector and dynamic validator, respectively.

Specifically, the first phase of {\name} (i.e., static detector) has reported 9,856 suspicious smart contracts, as they have shown code-level patterns of the vulnerability. 
As {\name} only reports 7,735 vulnerable contracts after the second phase (i.e., dynamic validator), there remains a gap of 2,121 contracts between these two phases.
\textit{Note that, there are no false positives introduced by {\name}, as all the 7,735 reported smart contracts can be exploited successfully.}

\subsubsection{Manually Investigation}
Next, we seek to analyze the 2,121 unreported ones, to see (1) how many of them are false positives introduced by \textit{static detector}, but later removed by our \textit{dynamic validator}, and (2) how many of them are false negatives of {\name}, which are indeed vulnerable ones but missed by the \textit{dynamic validator}. 
Furthermore, we also sample smart contracts from the unreported ones in the 176K tokens to evaluate (3) the overall coverage of {\name} (as the coverage is relying on the performance of static detector).

For the 2,121 smart contracts, 671 of them are open-source. 
Thus, we decide to select contracts from open-source ones for achieving accurate manual investigation.
We firstly sorted them by the lexicographic order based on the contracts addresses, and then grouped them into 10 subsets (67 contracts in first nine groups, and 68 contracts in the last one). 
Next, we chose 10 contracts randomly from each subset and analyzed them (overall 100 contracts) manually (\textit{Dataset-1}).
Moreover, we randomly select 100 smart contracts (with source code) from the unreported ones in the 176K ERC-20 tokens (\textit{Dataset-2}).
In total, we manually investigate 200 smart contracts.

\subsubsection{Effectiveness of {\name}} 
After two rounds of thorough and careful inspections by the first two authors, we have identified 90 false positive cases introduced by the static detector (but eliminated by dynamic validator) and 10 false negative cases from the 100 sampled smart contracts in \textbf{Dataset-1}. As to the \textbf{Dataset-2}, all of them are confirmed to be true negatives (i.e., without vulnerability).
This result suggests the effectiveness of {\name}, i.e., most of the vulnerable smart contracts can be automatically exploited by {\name}. 
\textbf{Overall, the precision of {\name} is 100\% (as all the reported cases are vulnerable), and the recall of {\name} is roughly over 99\%\footnote{It is estimated based on the proportion of false negative cases in Dataset-1 and Dataset-2, as roughly 212 cases (10\% * 2121) would be considered to be false negatives, among all the 176K contracts. 
Precision = TP/(TP+FP) and recall = TP/(TP+FN).}.
}

\subsubsection{False Positives of Static Detector}
The 90 false positives of static detector fall into three categories, including
\textit{inter-contract SafeMath} (72), \textit{stringent \texttt{throw} check} (15), and \textit{non-standard \texttt{transfer} implementation} (3).

\textbf{Inter-Contract SafeMath.}
As shown in Listing~\ref{lst:transfer-example}, some contracts implement \texttt{safeAdd}-like arithmetic operations in \texttt{SafeMath} contract, in order to perform an overflow verification by throwing an exception. 
However, the \texttt{SafeMath} is often implemented in another contract, and the token contract will inherit it and invoke the functions in it to perform arithmetic operations. For example, the \texttt{transfer} in listing~\ref{lst:transfer-example} calls two functions, i.e., \texttt{safeAdd} and \texttt{safeSub}, to prevent overflows. 
Nevertheless, on the bytecode level, the opcode sequence of \texttt{safeAdd} (and \texttt{safeSub}) can be either inlined into the \texttt{transfer} function, or invoked by the opcode \texttt{CALL}. In the former case, we can easily locate the \texttt{revert} opcode as shown in \S\ref{sec:approach:static}. However, in the latter case, because {\name} does not achieve the inter-contract symbolic execution (due to the limitation of Mythril), we cannot locate the \texttt{revert} operation and conduct the backward tracing accordingly, which will result in false positives.

\textbf{Stringent \texttt{throw} Check.}
As we mentioned in \S\ref{sec:approach:static:locate} and \S\ref{sec:approach:static:throw}, in the \texttt{transfer} (or \texttt{transferFrom}) function, we would firstly locate the variables being updated, and then trace backwards to verify if they are protected by the \texttt{throw} statement.
However, the semantic information of the bytecode is usually insufficient. As shown in the bottom right node in Fig.~\ref{fig:source-bytecode}, we are unable to distinguish the targets of \texttt{SSTORE} between \texttt{msg.sender} and \texttt{\_to}. Therefore, to guarantee the soundness of our static detector, we would check the verification on balance table for both of them, which inevitably introduces false positives as only the check for \texttt{balanceOf[msg.sender]} is necessary. 

\textbf{Non-Standard \texttt{transfer} Implementation.}
Even if the \texttt{transfer} function is implemented with the correct logic, it may still be implemented in a way that does not conform to the ERC-20 specification.
To be specific, as shown in listing~\ref{lst:wrong-transfer}, the \texttt{transfer} invokes \texttt{transferFrom} to complete the logic of transferring tokens.
Even if such an implementation could check the adequacy of balance of the invoker, this would not be consistent with the specification set out in ERC-20 guidance~\cite{erc20-guide} in which it strictly limits the behaviors of \texttt{transfer}, i.e., throwing an exception directly inside \texttt{transfer} if there is an insufficient balance.

\begin{lstlisting}[caption={An example of a mis-implementation of \texttt{transfer}.}, label={lst:wrong-transfer}]
function transfer(address dst, uint val) public returns (bool) {
  return transferFrom(msg.sender, dst, val);
}

function transferFrom(address src, address dst, uint val) public returns (bool)
{
  require(balanceOf[src] >= val);
  ...
}
\end{lstlisting}

\subsubsection{False Negatives of {\name}}
We have observed two reasons leading to the false negatives of {\name}: \textit{insufficient initialization} (9) and \textit{insufficient supply} (1).
Notice that, all of them are related to Type-II attack (see \S\ref{sec:vul:flawed-cex}).

\textbf{Insufficient Initialization.}
Our dynamic validator follows certain attack pattern that is described in \S\ref{sec:approach:dynamic} to validate the vulnerability. 
However, some tokes implement several non-standard requirements such that they need additional initialization. For example, \textit{BigAppleToken}~\cite{bigapple}, which implements all the mandatory functions in ERC-20 interface, has one special function as shown in listing~\ref{lst:bigappletoken}.
\begin{lstlisting}[caption={The code snippet in BigAppleToken.}, label={lst:bigappletoken}]
bool public transferEnabled;
function enableTransfer(bool _enable) external onlyOwner {
  transferEnabled = _enable;
  ... // following logic
}
\end{lstlisting}
Apart from that, in the \texttt{transfer} function, there is one more line (i.e., \texttt{require(transferEnabled)}) before all the statements, which can be regarded as a switch to enable the function.
Moreover, as the function names vary, it is impossible to take every situation into consideration. Therefore, such kind of tokens, which require additional initialization process, cannot pass the validator.

\textbf{Insufficient Supply.}
Under most circumstances, when the variable \texttt{totalSupply} of a contract is initialized, all the balance will be transferred to the owner's account, i.e., \texttt{msg.sender}.
However, in some cases, the initial supply is given to a designated address, like the code snippet shown in Listing~\ref{lst:insufficient-supply}. As it is impossible to assign the addresses of new created accounts even in the testnet of Ethereum, we cannot execute the following logic of dynamic validator unless modifying the source code.
\begin{lstlisting}[caption={An example of the insufficient supply problem.}, label={lst:insufficient-supply}]
address public founder = 0xCB7E...; // a fixed address
balances[founder] = totalSupply;
\end{lstlisting}
Moreover, in some cases, there might not exist initial supply at all. Thus we cannot perform the validation shown in Fig.~\ref{fig:fake-deposit-attack}(b).

\subsection{RQ3: The Impact of the Fake Deposit Vulnerability} 
We further measured the impact of the 7,735 influenced tokens. As described in RQ1, 56 ERC-20 tokens are affected by Type-I attack. 
The top-5 tokens (Type-I) with the highest number of holders and transactions are shown in Table~\ref{table:top5-vul-token} (Column 1-3).
They have 258 holders and 1,178 transactions in total.
Although the overall volume of them is not large, according to the validation methods we described in \S\ref{sec:approach:dynamic}, these 56 contracts can be attacked in the wild as long as the DEXes allow these tokens to be traded and the \texttt{depositToken} is used.
Consequently, we have identified three such kinds of DEXes, i.e., IDEX~\cite{idex}, DDEX~\cite{ddex}, and Ether Delta~\cite{etherdelata}. 
IDEX and DDEX are still active with a relatively high trading volume, around 1 million USD per day according to the statistics from \textit{etherscan.io}.

\begin{table}[]
\centering
\caption{Top 5 vulnerable tokens with most holders (Type-I) and largest market cap (Type-II).}
\vspace{-0.1in}
\resizebox{\columnwidth}{!}{%
\begin{tabular}{ccc|cccc}
\hline
\multicolumn{3}{c|}{\textbf{Type-I}}                 & \multicolumn{4}{c}{\textbf{Type-II}}                                     \\ 
\textbf{Token} & \textbf{\#Holders} & \textbf{\#Txs} & \textbf{Token} & \textbf{Cap(\$)} & \textbf{\#Holders} & \textbf{\#Txs}  \\ \hline\hline
CLB            & 53                 & 62             & BRC            & 391K             & 87K                & 1,382K          \\ \hline
BB             & 50                 & 32             & BAT            & 388K             & 305K               & 2,054K          \\ \hline
LOVE           & 49                 & 85             & HPT            & 63K              & 1K                 & 7K              \\ \hline
eDOGE          & 21                 & 21             & RPL            & 39K              & 3K                 & 30K             \\ \hline
EMVC           & 13                 & 15             & POWR           & 28K              & 50K                & 378K            \\ \hline\hline
\textbf{Total(56)} & \textbf{258}       & \textbf{1,178}   & \textbf{Total(7,716)} & \textbf{1.1B}    & \textbf{695K}      & \textbf{4.6M} \\ \hline
\end{tabular}%
}
\label{table:top5-vul-token}
\end{table}

For the 7,716 tokens that are vulnerable to Type-II attack, they are potentially at risk of being attacked in the scenario of insufficient validation of CEXes. 
Table~\ref{table:top5-vul-token} shows the top-5 tokens with the highest market capitalization.
Note that, here we consider fully diluted market cap, which is calculated by multiplying the token total supply with the current market price per token. The data is acquired on 10th June from \textit{etherscan.io}. 
If we take all the tokens suffered from Type-II attack into consideration, the market cap would be over 1 billion USD, and the number of holders and transactions would be 695K and 4.6 million respectively.
Therefore, if a CEX allows these tokens to be traded without comprehensive verification, the financial loss will be tremendous. 

%% file: discussion.tex
\section{Discussion}
\label{sec:discussion}

\noindent \textbf{Mitigation of Fake Deposit Vulnerability.}
For developers, strictly following the official guideline~\cite{erc20-guide} is always a good choice. 
To be specific, developers should implement all the six mandatory functions carefully, especially the \texttt{transfer} and \texttt{transferFrom} that are closely related to the transferring behaviors. The developer should \textit{throw} an exception if any mis-behavior happens (like insufficient balance or allowance) in both of them instead of returning a boolean value.
For the DEX, as both of the \texttt{transfer} and \texttt{transferFrom} will return a boolean value to indicate if the transferring succeeds, the DEX should never assume the return value from tokens. Additionally, it should handle the exception properly if the token raises it.
For the CEX, it should perform a comprehensive verification on each transaction of deposit request in back-end servers. 

\noindent \textbf{Limitation.}
We characterize the fake deposit vulnerability mainly from the perspective of the ERC-20 smart contracts. As we mentioned, performing the Type-II attack also requires the flawed verification of CEXes. However, we are unable to get the source code of CEXes. Thus, we only measure the scale of the vulnerable ERC-20 tokens, which is an upper bound for the Type-II attack.
Besides, we did not analyze the transactions on Ethereum, which may reveal the existence of real-world fake deposit attacks. 
Furthermore, it was reported that the fake deposit attacks have been observed in other blockchains (e.g., USDT~\cite{usdt-fake-deposit}, EOSIO~\cite{eosio-fake-deposit}), while this paper only focused on Ethereum.
We leave them for the future work.

%% file: related.tex
\section{Related Work}
\label{sec:related}
\noindent\textbf{Characterizing the Blockchain Ecosystem.} A number of studies have measured the blockchain ecosystem~\cite{huang2020characterizing,he2019characterizing,saad2019toward,xia2020characterizing,jourdan2018characterizing,chen2018understanding}. Chen et al.~\cite{chen2018understanding} characterized money transfer, contract creation and invocation of Ethereum through graph analysis. Huang et al.~\cite{huang2020characterizing} characterized EOSIO blockchain, and \cite{jourdan2018characterizing} proposed an approach to identify entities in Bitcoin blockchain.

\noindent\textbf{Program Analysis of Smart Contract.} Based on program analysis techniques, e.g., symbolic execution and formal verification, a variety of frameworks have been proposed to improve the security of smart contracts~\cite{grech2018madmax,kalra2018zeus,brent2018vandal,tikhomirov2018smartcheck,feist2019slither,mossberg2019manticore,tsankov2018securify,he2020security}. Specifically, M. Mossberg et al.~\cite{mossberg2019manticore} proposed Manticore, a dynamic symbolic execution framework for analyzing Ethereum smart contract. He et al.~\cite{he2020security} implemented a symbolic execution framework for analyzing smart contracts at WebAssembly level, which is used by EOSIO blockchain.

\noindent\textbf{ERC-20 Tokens.}
Several work~\cite{fenu2018ico,victor2019measuring,somin2018network,frowis2019detecting,somin2020erc20,chen2020traveling,chen2019tokenscope} have focused on the ERC-20 smart contracts in Ethereum. A number of studies~\cite{victor2019measuring,somin2018network,chen2020traveling} analyzed the network structures of tokens and transactions in Ethereum. Chen et al. proposed TokenScope~\cite{chen2019tokenscope}, which is able to detect transactions that triggers inconsistent behaviors. 
Somin et al.~\cite{somin2020erc20} presents the analysis of the dynamical properties of the ERC-20 protocol, which in turn allows the prediction of some network parameters.

%% file: conclusion.tex
\section{Conclusion}
\label{sec:conclusion}

In this work, we have systematically characterized the fake deposit vulnerability in Ethereum. 
{\name}, an automated tool is proposed to perform the detection and verification of the vulnerability.
We demonstrate the efficiency of {\name} with experiments on a large number of smart contracts.
Our observations reveal the prevalence of fake deposit vulnerability in the ERC-20 smart contracts.
Our efforts can positively contribute to bring developer awareness,
attract the focus of the research community and regulators,
and promote best operational practices across blockchains.